\documentclass{osa-article}

\journal{osac}

\articletype{Research Article}

\begin{document}

\title{Highly efficient cooling of mechanical resonator with square pulse drives}

\author{Qing Lin\authormark{1,3} and Bing He\authormark{2,4}}

\address{\authormark{1}Fujian Key Laboratory of Light Propagation and Transformation, College of Information Science and Engineering, Huaqiao University, Xiamen 361021, China\\
\authormark{2}Center for Quantum Optics and Quantum Information,
Universidad Mayor, Camino La Pir\'{a}mide 5750, Huechuraba, Chile\\
\authormark{3}qlin@hqu.edu.cn\\
\authormark{4}binghe9988@gmail.com}



\begin{abstract}
Ground state cooling of mechanical resonator is a way to generate macroscopic quantum states. Here we present a study of optomechanical cooling under the drive of square pulses without smooth profile. By illustrating the dynamical processes of cooling, we show how to choose 
the amplitudes and durations of square pulses, as well as the intervals between them, so that a mechanical resonator can be quickly cooled down to its ground state. Compared with the cooling under a continuous-wave drive field, the ground state cooling of a mechanical resonator can be performed more efficiently and flexibly by using square pulse drives. At certain times of such cooling process, the thermal phonon number under square pulse drives can become even lower than the theoretical limit for the cooling with a continuous-wave drive field of the same amplitude.
\end{abstract}

\section{Introduction}
Over a century from the emergence of quantum physics, people keep wondering whether or not there exists a boundary between classical and quantum world \cite{decoherence,boundary1,boundary2}. Finding this boundary is a fundamental issue to quantum mechanics, which implies that quantum states can exist even for macroscopic objects. The possible creation of macroscopic quantum objects has, therefore, attracted tremendous interests. However, the 
ubiquitous thermal noise from environment always decoheres macroscopic quantum states, making it hard to see macroscopic quantumness \cite{decoherence}. 
The attempts to create macroscopic quantum
states in thermal environment have been made with the platform of optomechanical systems (OMS) \cite{OMS1, qnano}, including both theoretical proposals (see, e.g. \cite{coolt1,coolt2,coolt3,coolt4,coolt5,coolt6,coolt7,coolt8,coolt9,coolt10,coolt11,coolt12,coolt13}) and experimental realizations (see, e.g. \cite{coole1,coole2,coole3,coole4,coole5,coole6,coole7,coole8,coole9,coole10,coole11,coole12,coole13,coole14}). 

So far the study of OMSs has been extended to the setups that apply the drives in the form of pulses \cite{pulsed1,pulsed2}. 
In 2011 Hofer et al. investigated the entanglement creation with pulsed OMS entanglement \cite{pulsetheo1}, and the similar systems have been experimentally realized to implement mechanical resonator cooling \cite{pulseex1, pulseex2}. Like the other treatments of pulsed OMSs \cite{Liao, Yli}, 
the theoretical description of the dynamical processes of an OMS under pulsed drive is based on the linearization by expanding the cavity mode operator around its classical mean value $\alpha(t)$, a straightforward generalization of a similar approach to OMSs under continuous-wave (CW) drive \cite{OMS1, qnano}. The time-dependent cavity mean field  $\alpha(t)$ can be determined with good approximation for a drive field $E(t)$ varying slowly over a time scale $1/\kappa$ (the inverse of the damping rate of a cavity) \cite{pulsetheo1}. The validity of such method thus requires that the used pulses should have smooth heads and tails. There are, however, many kinds of pulses that do not have smooth profiles. Square pulse, which is commonly used in optics labs, is a typical example of this type. The previous theoretical approaches to pulsed OMSs are not applicable to square pulses, which abruptly change their amplitudes at their heads and tails.

In the current work we fill this gap by applying a different approach to the problem of optomechanical cooling under pulse drives. 
The different approach \cite{cooling} we use enables one to find the details of the 
dynamical evolutions in cooling processes, for arbitrary pulse intensity, duration and interval between pulses. We find that a mechanical resonator can be efficiently cooled down to its ground state by a series of square pulses that are properly designed. Furthermore, a ground state cooling can be achieved quickly and well preserved by adjusting the intensity and duration of square pulses, as well as the interval between pulses. A rather interesting feature of cooling with square pulse is that, at certain times of a cooling process, it can even achieve better cooling result than the corresponding 
theoretical cooling limit for CW drive field.

The rest of the paper is organized as follows. In Sec. 2, the effective OMS Hamiltonian and the corresponding dynamical equations for a cooling process with pulse are derived. The cooling by a single pulse drive field is discussed in Sec. 3, and the cooling under a series of identical pulses with modulated intensity is illustrated by the examples in Sec. 4, associated with the relations between the cooling rate, the cooling speed and the effective intensity for achieving fast ground state cooling. The improvement of the cooling with a series of strong pulses 
designed according to those relations is discussed in Sec. 5. In Sec. 6, we present a comparison of the mechanical motion predicted with our approach with those numerically simulated with the full dynamical equations without linearization, to provide an evidence for the validity of our approach. Finally this work is concluded in Sec. 7.

\section{System Hamiltonian and dynamical equations}
\label{sec2}

\begin{figure}[b]
\centering\includegraphics[width=9cm]{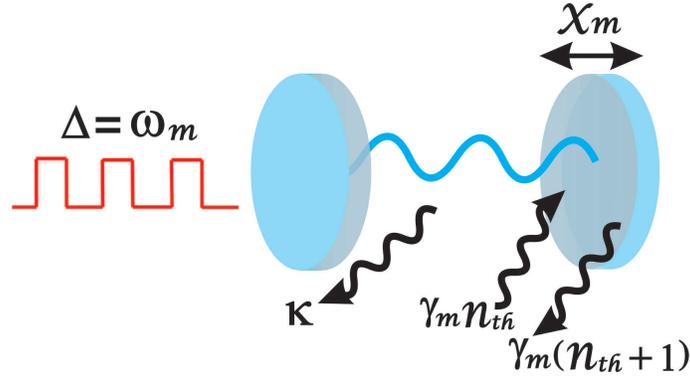}
\caption{Model of pulsed optomechanical cooling. The pulse drive field with red detuning $\Delta=\omega_m$ is sent into the cavity with the one movable mirror (mechanical resonator), which will be cooled down to its ground state given a proper profile of the pulses.}  \label{pulse}
\end{figure}
We consider an OMS in Fig. \ref{pulse}, which is driven by a series of pulsed laser drive with the central 
frequency $\omega_L$ and a square profile of the amplitude $E$ and the duration $t_0$. The Hamiltonian of the system reads ($\hbar=1$),
\begin{align}
H(t)=&\underbrace{\omega_c \hat{a}^{\dag}\hat{a}+\omega_m  \hat{b}^{\dag}\hat{b}+i[\hat{a}^{\dag}E(t)e^{-i\omega_L t}-\hat{a}E^*(t)e^{i\omega_L t}]}_{H_S(t)}\underbrace{-g_{m}\hat{a}^{\dag}\hat{a}(\hat{b}+\hat{b}^{\dag})}_{H_{OM}} \nonumber\\
&+\underbrace{i\sqrt{2\kappa}\{\hat{a}^{\dag} \hat{\xi}_c(t)-\hat{a}\hat{\xi}^{\dag}_c(t)\}+i\sqrt{2\gamma_m}\{\hat{b}^{\dag} \hat{\xi}_m(t)-\hat{b}\hat{\xi}^{\dag}_m(t)\}}_{H_{SR}}.
\end{align}
The first part $H_S(t)$ involves the free Hamiltonians of the cavity mode and the mechanical mode with the resonant frequency $\omega_c$ and $\omega_m$, respectively, together with the external drive. The second part $H_{OM}$ is the coupling of the cavity mode with the mechanical mode due to the radiation pressure, where $g_m$ is the single-photon coupling strength. The couplings of the cavity and mechanical modes with the environmental reservoirs are described by the linear coupling term $H_{SR}$. The corresponding stochastic Langevin noise operator $\hat{\xi}_c$ ($\hat{\xi}_m$) of the reservoir satisfies the relation $<\hat{\xi}^\dagger_l(t)\hat{\xi}_l(\tau)>_R=n_l\delta(t-\tau)$ ($l=c,m$) with the occupation number $n_l$ in thermal equilibrium. 
Before a cooling procedure one has $n_c=1/(e^{\hbar \omega_c/k_B T}-1)\approx 0$ and $n_m=1/(e^{\hbar \omega_m/k_B T}-1)$ that is also denoted as the thermal occupation $n_{th}$.

The evolution of such OMS can be described by the generalized evolution operator $U(t)=\mathcal{T}\exp\{-i\int_0^t d\tau \hat{H}(\tau)\}$ as a time-ordered exponential \cite{book}. The nonlinear term $H_{OM}$ makes it hard to find the analytical solution for the evolution operator $U(t)$. 
Here, we adopt the approach developed in \cite{cooling} to the problem. By applying the dynamical approach in \cite{cooling}, it is not necessary to find the mean values of the system's dynamical variables (such as $\langle \hat{a}(t)\rangle$) that must be obtained by solving the corresponding classical nonlinear equations of OMS. This flexibility enables one to work without the mean cavity field $\alpha(t)=\langle \hat{a}(t)\rangle$ due to a square pulse, which is 
impossible to find with high precision by the previous approach to pulsed OMS \cite{pulsetheo1}. The similar dynamical approaches to the one in \cite{cooling} have been applied to some other physical systems \cite{He,Lin1,rydberg, Lin2,noise, dynamics, pulse1, Lin3} too.

We first apply a decomposition of the evolution operator as follows:
\begin{align}
U(t)=&\mathcal{T}\exp\{-i\int_0^t d\tau H_S(\tau)\} \times\mathcal{T}\exp\{-i\int_0^t d\tau [H_{eff}(\tau)+H_N(\tau)]\},
\end{align} 
where $H_{eff}(\tau)+H_N(\tau)=\mathcal{T}\exp\{i\int_0^\tau dt' H_S(t')\}\{ H_{OM}+H_{SR}(\tau)\}\mathcal{T}\exp\{-i\int_0^\tau dt' H_S(t')\}$. 
The forms of the cavity and mechanical operators in the effective Hamiltonian $H_{eff}(\tau)$ and $H_N(\tau)$ are found as
\begin{align}
&\mathcal{T}\exp\{i\int_0^t d\tau \hat{H}_S(\tau)\}~\hat{a}~\mathcal{T}\exp\{-i\int_0^t d\tau \hat{H}_S(\tau)\}\nonumber\\
&=e^{-i\omega_c t}(\hat{a}+\int_0^t d\tau E(\tau)e^{i\Delta \cdot\tau})\equiv e^{-i\omega_c t}(\hat{a}+F(t)),\label{trf}\\
&\mathcal{T}\exp\{i\int_0^t d\tau \hat{H}_S(\tau)\}~\hat{b}~\mathcal{T}\exp\{-i\int_0^t d\tau \hat{H}_S(\tau)\}=e^{-i\omega_mt}\hat{b}, 
\end{align}
where the detuning is $\Delta=\omega_c-\omega_L$ and $F(t)$ is a time-dependent function. Therefore, we have
\begin{align}
H_{eff}(\tau)=&-g_m\left[F(\tau)\hat{a}^{\dag}+F^*(\tau)\hat{a}+|E(\tau)|^2\right]\left(e^{-i\omega_m\tau}\hat{b}+e^{i\omega_m\tau}\hat{b}^{\dag}\right) \nonumber\\
&+i\sqrt{2\kappa}\left\{e^{i\omega_c \tau}\left(\hat{a}^{\dag}+F^*(\tau)\right) \hat{\xi}_c(\tau)-e^{-i\omega_c \tau}\left(\hat{a}+F(\tau)\right)\hat{\xi}^{\dag}_c(\tau)\right\}\nonumber\\
&+i\sqrt{2\gamma_m}\left(e^{i\omega_m\tau}\hat{b}^{\dag}\hat{\xi}_m(\tau)-e^{-i\omega_m\tau}\hat{b}\hat{\xi}^{\dag}_m(\tau)\right),\\
H_N(\tau)=&-g_m\hat{a}^{\dag}\hat{a}\left(e^{-i\omega_m\tau}\hat{b}+e^{i\omega_m\tau}\hat{b}^{\dag}\right).
\end{align} 
The linearized equations of motion due to $H_{eff}(t)$ are the following: 
\begin{align}
\dot{\hat{a}}=&-\kappa \hat{a}+i g_m F(t) (e^{-i\omega_mt}\hat{b}+e^{i\omega_mt}\hat{b}^{\dag})-\kappa F(t)+\sqrt{2\kappa}e^{i\omega_c t}\hat{\xi}_c (t),\nonumber\\
\dot{\hat{b}}=&-\gamma_m \hat{b}+i g_m e^{i\omega_m t}\left[F(t)\hat{a}^{\dag}+F^*(t)\hat{a}\right]+i g_m e^{i\omega_m t}|F(t)|^2+\sqrt{2\gamma_m}e^{i\omega_mt}\hat{\xi}_m(t), \label{dye}
\end{align}
where the effect of $H_N(\tau)$ is neglected under the condition $g_m/\omega_m \ll 1$ \cite{cooling}. Two terms proportional to $F(t)$ ($F^*(t)$) exist in each of the 
above equations; the one containing the operator $\hat{a}$ ($\hat{b}$) indicates the effect of swapping between the cavity and mechanical mode, while the other with $\hat{a}^\dagger$ ($\hat{b}^\dagger$) manifests a squeezing effect on the two modes.
In what follows, we will numerically simulate the evolutions of thermal phonon 
number 
\begin{align}
n_m(t)
=& \left\langle \underbrace{(\hat{b}^{\dag}(t)-\langle\hat{b}^{\dag}(t)\rangle)}_{\delta \hat{b}^{\dag}(t)}\underbrace{(\hat{b}(t)-\langle\hat{b}(t)\rangle)}_{\delta \hat{b}(t)}\right\rangle \nonumber\\
=&\langle\hat{b}^{\dag}\hat{b}(t)\rangle-\langle\hat{b}^{\dag}(t)\rangle\langle\hat{b}(t)\rangle
\end{align}
defined with the mechanical fluctuation $\delta \hat{b}(t)$. In this definition $n_m(t)$ we exclude the quantity $\langle\hat{b}^{\dag}(t)\rangle\langle\hat{b}(t)\rangle$ due to the kinetic motion of mechanical resonator, which is generally not zero under pulsed drives. At $t=0$ when the mechanical resonator is in a thermal state (thermal equilibrium with the environment),
there is $\langle\hat{b}\rangle=0$ so that $n_m(0)=n_{th}$.

\section{Cooling under a single pulse}
\label{sec3}

\begin{figure}[t!]
\centering\includegraphics[width=9cm]{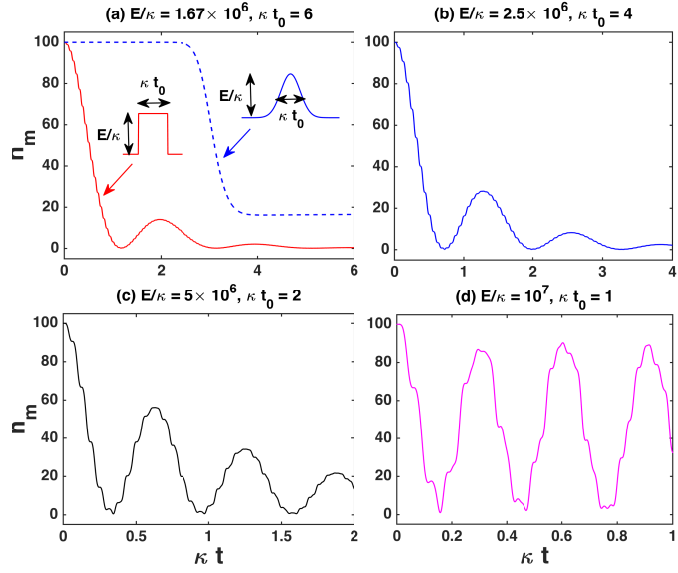}
\caption{Evolution of phonon number under a single square pulse drive field with different drive intensity $E/\kappa$ and duration $\kappa t_0$, together with that under a single Gaussian pulse drive field. For comparison, 
the product $(E/\kappa)\times(\kappa t_0)$ is set to be fixed as $10^7$. The mechanical resonator is seen to be quickly cooled down with strong intensity, but the final phonon number will oscillate due to a stronger pulse amplitude. The other parameters are $g_m/\kappa=10^{-4}$, $\Delta/\kappa=\omega_m/\kappa=100$, $\gamma_m/\kappa=10^{-3}$, and $n_{th}=100$.}  \label{sgd}
\end{figure}

The dynamical equations in Eq. (\ref{dye}) can be actually applied to study the cooling under a pulse of arbitrary profile $E(t)$. If a single square pulse is applied, the function $F(t)$ in Eq. (\ref{trf}) takes the form 
\begin{eqnarray}
F(t)=&\left\{\begin{matrix}
 &iE/\Delta(1-e^{i\Delta t}); \text{  $t\leqslant t_0$}, \\ 
 &0; \text{  $t>t_0$}.
\end{matrix}\right.
\end{eqnarray}
The detuning is set to be $\Delta=\omega_m$, the resonance point to achieve the maximum beam splitter effect, which leads to the cooling of the mechanical resonator. Besides this detuning, the intensity $E$ and the duration $t_0$ of the square pulse are two other crucial parameters. With a fixed product $E\cdot t_0=10^7$, we display the results for four cases, $\kappa t_0=6, 4, 2, 1$ in Fig. \ref{sgd}, with the initial phonon number $n_m=100$ under the thermal equilibrium with the environment. For comparison, we also display an example of the evolution of phonon number under a Gaussian pulse with the profile $E(t-j_0)=Ee^{-\sigma^2(t-j_0)^2}$ ($E/\kappa=1.67\times10^7, \sigma/\kappa=1/3$). The mechanical resonator can hardly be cooled down to ground state with a Gaussian pulse with the comparable intensity and width, as the Gaussian one acts less strongly with its uneven amplitude. One also see the jagged thermal phonon number curves 
following various tendencies under different conditions, and it comes from the small oscillations of the quantity with the frequency $\omega_m$. 

From the results in Fig. \ref{sgd}, we find two features in such cooling:

1. The intensity $E$ determines the cooling speed (how soon the first dip of thermal phonon number is reached). 
Since the narrower the pulse is, the stronger of the pulse will be under the condition $E\cdot t_0=10^7$, a faster cooling of the mechanical resonator is possible with a narrower pulse. For instance, the phonon number $n_m$ is quickly reduced from $100$ to $0.65$ within the dimensionless time $\kappa t\sim 0.34$ in Fig. \ref{sgd}(c). The increase of the intensity will initially reduce the time to the reach the first minimal value of the thermal phonon number.

2. The intensity $E$ also determines the oscillation pattern of the evolved thermal phonon number as in Fig. \ref{sgd}. If the drive intensity becomes still larger, the evolved phonon number will finally oscillate between two values of large difference, as the one between $1$ to $90$ in Fig. \ref{sgd}(d). The competition between the comparable cooling effect (the swapping between the cavity and mechanical mode) and heating effect due to the remnant two-mode squeezing effect as shown in Eq. (\ref{dye}) leads to such oscillation pattern.

Therefore, a high pulse amplitude helps to improve the cooling speed, but the mechanical resonator cannot be efficiently cooled down well if 
the pulse is too strong as in Figs. \ref{sgd}(c) and \ref{sgd}(d). A ground state cooling can be realized with proper pulse amplitude. For example, the approximate steady phonon number $n_m=0.14$ is realized under $E/\kappa=1.67\times10^6$ as in Fig. \ref{sgd}(a). There is a trade-off between 
the cooling speed and final phonon number, which is similar to the one discovered for the cooling with a CW drive field \cite{cooling}.

\section{Cooling under a series of identical pulses}
\label{sec4}

\begin{figure*}[t]
\centering\includegraphics[width=13.5cm]{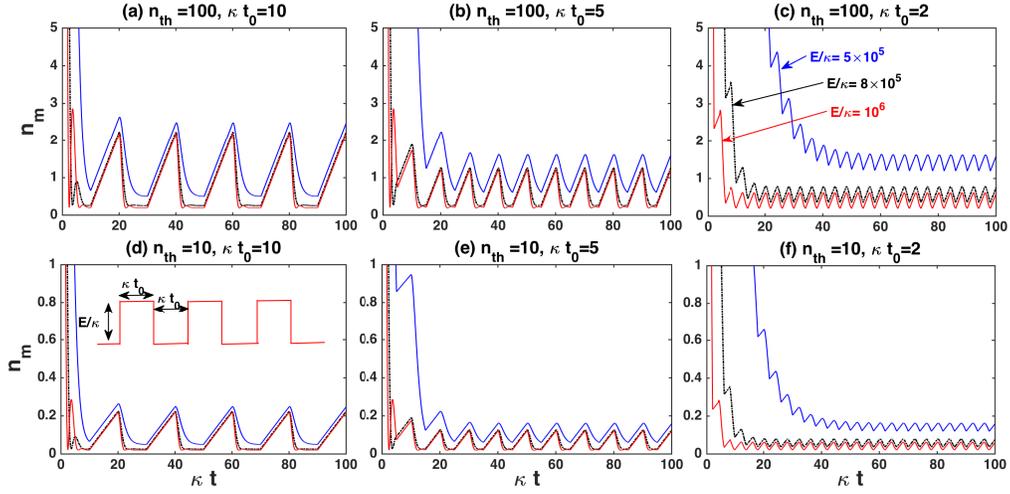}
\caption{Evolution of phonon number under multiple pulses. The duration of and interval between the pulses are set to be equal. The ground state cooling can be achieved and the ground state can be preserved with the duration $\kappa t_0=2$ and the intensities $E/\kappa=8\times10^5, 10^6$, while the ground state cannot be well preserved given 
the longer durations $\kappa t_0=10, 5$ or cannot be achieved given the relatively weak intensity $E/\kappa=5\times10^5$. The other parameters are chosen as $g_m/\kappa=10^{-4}$, $\Delta/\kappa=\omega_m/\kappa=100$, 
and $\gamma_m/\kappa=10^{-3}$.}  \label{md}
\end{figure*}

Next we discuss the cooling under a series of identical square pulses. Here we assume that the duration and interval are $t_1$ and $t_2$, respectively, 
so that the function $F(t)$ takes the form 
\begin{align}
F(t)=&\left\{\begin{matrix}
 &\frac{iE}{\Delta}(1-e^{i\Delta (t-j(t_1+t_2))}); \text{  $0<t-j(t_1+t_2)\leqslant t_1$}, \\ 
 &0; \text{$0<t-j(t_1+t_2)-t_1\leqslant t_2$,}
\end{matrix} \right. 
\end{align}
where $j=\lfloor \frac{t}{t_1+t_2} \rfloor$ (taking the integer part of $\frac{t}{t_1+t_2}$) is the numbering for the pulses one after another. Here the detuning is fixed to be $\Delta=\omega_m$ too. The mechanical resonator can be cooled under the drive of one of the pulses, but will be heated up in the interval between two pulses because it is in contact with the thermal environment. During the interval with $F(t)=0$ the thermal phonon number for the mechanical resonator will increase by 
\begin{eqnarray}
\Delta n_m=2\gamma_m n_{th} \Delta t,
\end{eqnarray}
where $\Delta t=t-j(t_1+t_2)-t_1$. Another fact is that the simultaneous heating due to the remnant squeezing effect (when the pulsed drive is on) will also be significant, if the pulse drive intensity is high.

\begin{figure}[b]
\centering\includegraphics[width=9cm]{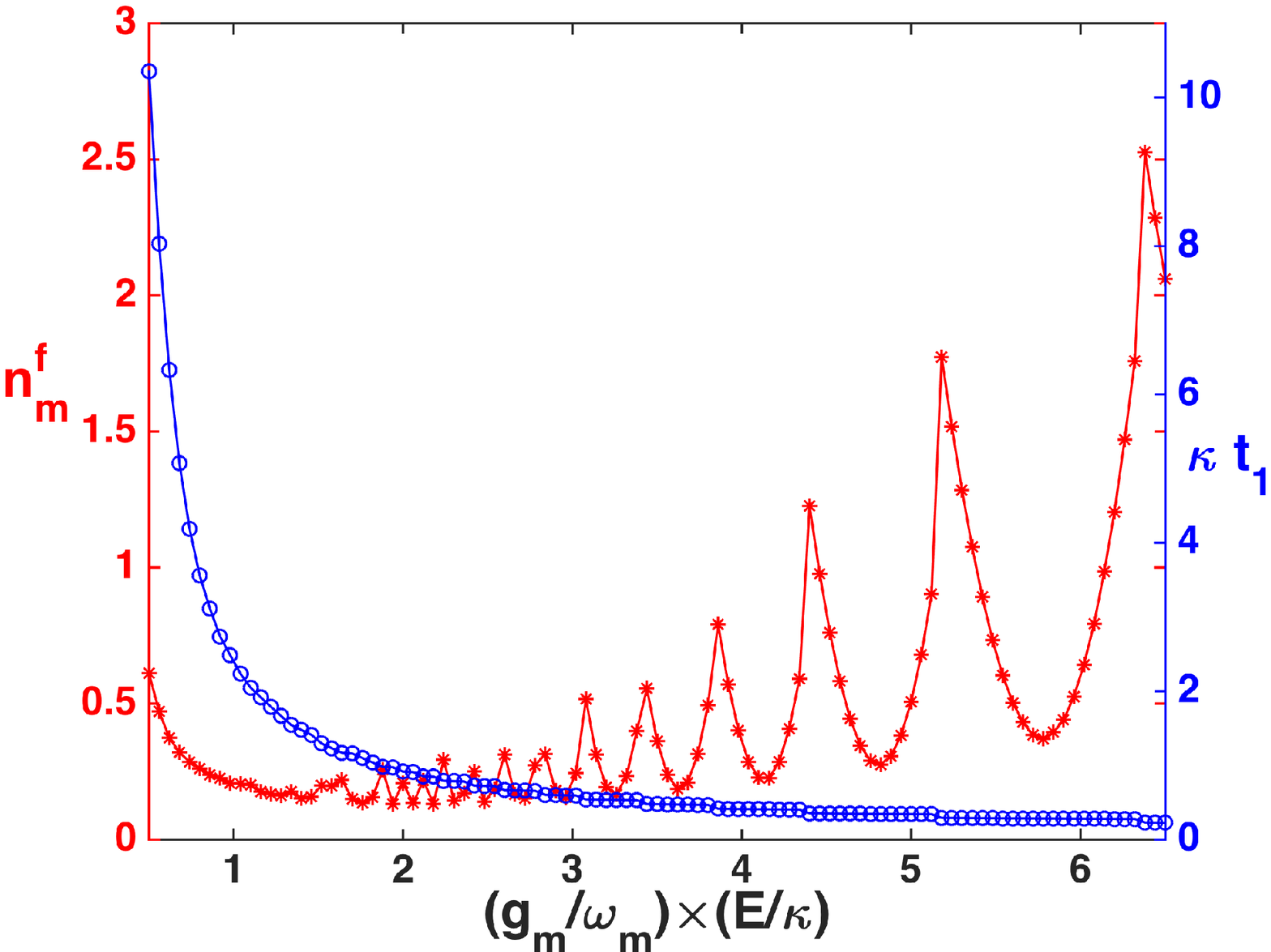}
\caption{Relations between the time of reaching the first dip (blue circle) and the corresponding phonon number (red star) with the dimensionless effective intensity $J=(g_m/\omega_m)\times(E/\kappa)$. Increasing the effective intensity will increase the cooling speed (a shorter time of first dip), but the corresponding phonon number will become oscillating with the parameter. The system parameters are set as $g_m/\kappa=10^{-4}$, $\Delta/\kappa=\omega_m/\kappa=100$, $\gamma_m/\kappa=10^{-3}$, and $n_{th}=100$.}  \label{speed}
\end{figure}

With the three different drive amplitudes $E/\kappa=5\times 10^5, 8\times 10^5$ or $10^6$ in Fig. \ref{md}, we illustrate how the thermal phonon number evolves given two initial values of $n_{th}=100$ or $10$. Here the pulse durations are set as $\kappa t_0=10$, $5$ or $2$. As shown in these figures, the mechanical resonator will be cooled down to the ground state, and the phonon number will finally become a steady value with a moderate pulse amplitude, just like the situation in Fig. \ref{md}(a). However, during the interval between two pulses the mechanical resonator will be heated up out of the ground state, 
if the interval is long enough. Such periodic change makes the phonon number of the mechanical resonator oscillate around a low value. 

When the pulse duration is quite shorter as in Fig. \ref{md}(b), the mechanical resonator cannot be immediately cooled to the ground state by the first pulse, but the action of the consequent pulses will achieve the purpose. In Fig. \ref{md}(c), the duration is very short as $\kappa t_0=2$. The mechanical resonator cannot be cooled to the ground state with a weaker drive intensity $E/\kappa=5\times 10^5$, to make the cooling speed slow. 
On the contrary, the mechanical resonator can be cooled down to the ground state after two or three pulses given the intensity $E/\kappa=10^6$ and $E/\kappa=8\times 10^5$. After that, the mechanical resonator can be well preserved in the ground state, since the drive intensity is strong enough.
Through the comparisons one finds that short duration, together with a relatively strong intensity, e.g., $E/\kappa=8\times 10^5$ or $10^6$, 
is good to preserve the mechanical resonator in the ground state, while a relatively long duration (e.g. $\kappa t_0=5$) can realize a ground state with 
low intensity.  

One can expect that a strong pulse drive makes the cooling fast. It is verified with a relation in Fig. \ref{speed}, which shows how the time (blue circles) for reaching the first minimal value of phonon number and the corresponding phonon number $n^f_m$ (red stars) change with a dimensionless parameter $J=(g_m/\omega_m)\times(E/\kappa)$. Obviously, the relation between the time at the first dip and the drive intensity is consistent with the one in Fig. \ref{sgd}. However, the relation between the phonon number $n^f_m$ and the pulse intensity is less trivial. When the intensity is not strong, 
the phonon number at the first dip decreases with the increased drive intensity. But it will oscillate with the intensity $E$ 
(all other parameters of the system are assumed to be fixed) when the intensity becomes relatively strong (as the parameter $J=(g_m/\omega_m)\times(E/\kappa)$ is larger than $1.5$ in Fig. \ref{speed}). 
This oscillation pattern reflects the existence of the competition between a squeezing effect (heating) and a beam splitter effect (cooling) as the drive intensity keeps increasing. These relations provide good guidance for selecting the proper parameters for cooling with square pulsed field. Finding the dynamical evolutions of the cooling processes is necessary for obtaining such relations, and it is possible only with a fully dynamical approach 
like that in Sec. II.    

By appearance, cooling with multiple square pulses is no much different from cooling with a CW drive. For example, if the drive intensity is increased, the cooling of the mechanical resonator can be accelerated but, meanwhile, the mechanical resonator will not be cooled down well due to the simultaneously enhanced heating from a remnant two-mode squeezing effect. Does it mean that a drive field with strong intensity is not suitable for cooling? In the situation of cooling with pulses, such restriction can be much relaxed by designing the profile of the used pulses properly. Taking this advantage, one can cool down a mechanical resonator to a ground state and preserve the ground state with a rather strong pulse intensity nonetheless.
The procedure to do so goes as follows:

1. To act a strong pulse, which is properly designed according to the relation in Fig. \ref{speed}, to the mechanical resonator;

2. To stop the action of the pulse at the time of reaching the first minimal value of phonon number before the simultaneous heating becomes significant (this step can be automatically performed by choosing the proper duration of and the interval between pulses); 

3. To keep the repetition of the above with a properly designed strong pulse series so that the mechanical resonator can be preserved in a ground state.

We use some examples to illustrate how it works. Here the drive intensity is set to be $E/\kappa=5\times10^6$, which is the same as that in Fig. \ref{sgd}(c) where the cooling with a single such pulse is not very effective. The dimensionless time $\kappa t$ for the pulse to reach the first dip of phonon number is at about $0.343$, so we set the pulse duration $\kappa t_1=0.34$. For the examples in Fig. \ref{optimal}, the time intervals between pulses are chosen as $\kappa t_2=0.34, 1.0, 1.5, 2.0, 2.5, 3.0$, respectively. Obviously, as one can see from the extension of Fig. \ref{sgd}(c) to longer time, the mechanical resonator can be hardly kept to be in a ground state, if the pulse drives are replaced by a CW one of the same intensity. 
\section{Advantage of cooling with square pulse}
\label{sec5}
\begin{figure*}[t!]
\centering\includegraphics[width=13.5cm]{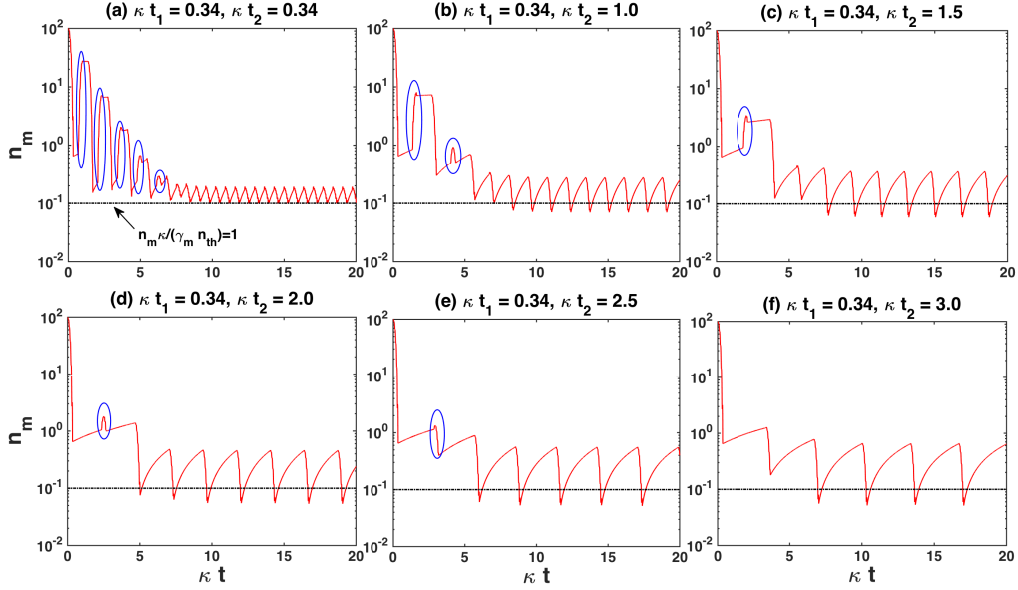}
\caption{Evolution of phonon number with the short duration $\kappa t_1=0.34$ and different time interval $\kappa t_2=0.34, 1.0, 1.5, 2.0, 2.5, 3.0$. 
The intensity is strong as $E/\kappa=5\times 10^6$. The dash-dotted line denotes the theoretical cooling limit $n_m\kappa/(\gamma_m n_{th})=1$ with continuous-wave drive field, which can be beyond by the pulsed optomechanical cooling. The ellipses mark the periods, in which the mechanical resonator is heated up when the pulses are applied. All other parameters are $g_m/\kappa=10^{-4}$, $\Delta/\kappa=\omega_m/\kappa=100$, $\gamma_m/\kappa=10^{-3}$, and $n_{th}=100$.}  \label{optimal}
\end{figure*}

Here the mechanical resonator can be quickly cooled to a ground state ($n_m <1$) under the first pulse. When the interval between pulses is short, such as $\kappa t_2=0.34, 1.0, 1.5, 2.0$, the mechanical resonator will be obviously heated afterward (the places marked with the ellipses). 
Especially in Fig. \ref{optimal}(a), the heating tendency exists even after the action of the tenth pulse. When the pulse interval is as long as $\kappa t_2=2.5, 3.0$, the heating tendency will become less obvious from the second pulse duration. It implies that the heating due to the thermal environment itself is much less significant than the heating from a simultaneously acting two-mode squeezing effect when a sufficiently strong pulse is on. Nevertheless, the existence of heating tendency will not affect the cooling in the end. For example, the mechanical resonator can be well preserved in ground state (the phonon number is about $0.1 \sim 0.2$) as in Fig. \ref{optimal}(a), after the system goes through five periods of being heated up again (the five periods marked with the ellipses).

An interesting feature is that the pulsed cooling can reach a minimum thermal phonon number that is below the limit $n_m\kappa/(\gamma_m n_{th})=1$ for cooling by a CW field \cite{cooling}, which is indicated as the horizontal dot-dashed lines in Figs. \ref{optimal}(a)-\ref{optimal}(d). During the evolution the phonon number periodically becomes even lower than the cooling limit value. This phenomenon exhibits another special character of cooling with pulsed drive field.

\section{Evidence for the validity of the present approach}
\label{sec6}
\begin{figure*}[t!]
\centering\includegraphics[width=13.45cm]{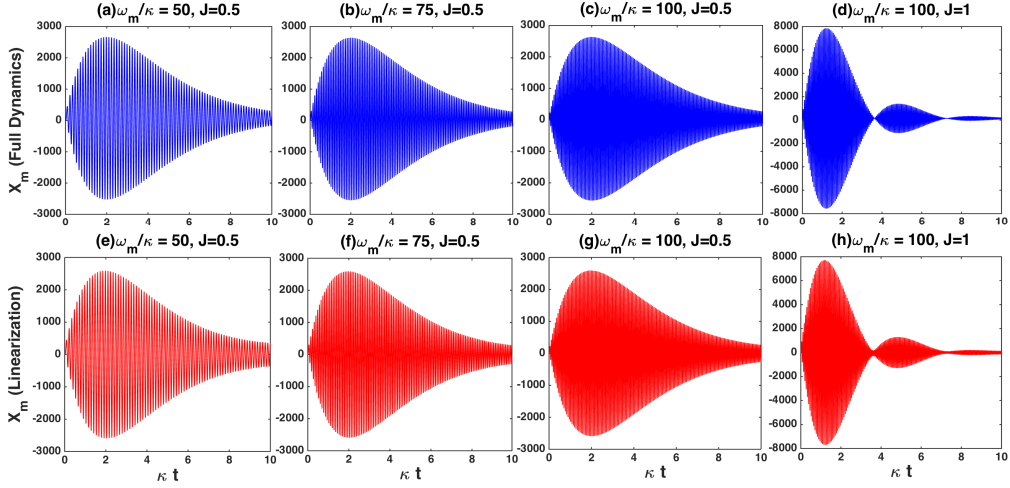}
\caption{Comparisons of the mechanical resonator's displacements $X_m(t)$ predicted with the nonlinear dynamical equations Eq. (\ref{neq}) (blue) 
and our linearized equations Eq. (\ref{dye}) (red). Theses comparisons are made with some different parameters $\omega_m/\kappa$ and $J=(g_m/\omega_m)\times(E/\kappa)$.
The common parameters for the different situations are chosen as $g_m/\kappa=10^{-4}$, $\Delta/\kappa=\omega_m/\kappa$, and $\gamma_m/\kappa=10^{-3}$.}  \label{figc200}
\end{figure*}

The cooling results presented in the paper are found with Eq. (\ref{dye}), the linearized equations of motion derived with the procedure in Sec. II.
In reality, a nonlinear term $H_{N}$ is contained in the effective Hamiltonian so that the exact Hamiltonian $H(t)$ for the OMS gives rise to 
the following nonlinear quantum Langevin equations: 
\begin{align}
\dot{\hat{a}}=&-\kappa \hat{a}+i g_m (e^{-i\omega_mt}\hat{b}+e^{i\omega_mt}\hat{b}^{\dag})\hat{a}+Ee^{i\Delta t}+\sqrt{2\kappa}\hat{\xi}_c (t)\nonumber\\
\dot{\hat{b}}=&-\gamma_m \hat{b}+i g_m e^{i\omega_m t}\hat{a}^{\dag}\hat{a}+\sqrt{2\gamma_m}\hat{\xi}_m(t). \label{neq}
\end{align}
The direct application of the above nonlinear equations to calculate cooling results is impossible thus far.
To measure how good our linearized equations reflect the actual cooling processes under pulsed drives, we compare the predictions of another dynamical quantity by the above nonlinear equations and with our linearized Eq. (\ref{dye}), respectively. It is natural to hold that dynamical quantities predicted by the two different sets of equations have one-to-one correspondence---if the predictions of one of the dynamical quantities match well, those for all other quantities (including the thermal phonon number) would have small difference.

Here we use the displacement $X_m(t)=\frac{1}{\sqrt{2}}\langle \hat{b}(t)+\hat{b}^\dagger (t)\rangle$ of the moving mechanical resonator as such quantity for indicating the closeness of the predictions by the full nonlinear equations and our linearized equations of motion. In the simulation with the nonlinear Eq. (\ref{neq}), we assume the factorization of the nonlinear factors \cite{eit}, e.g. $\langle \hat{b}^{\dag}\hat{a}\rangle=\langle \hat{b}^{\dag}\rangle \langle \hat{a}\rangle$. 
The comparisons of a group of mechanical displacements $X_m(t)$ predicted by the two different sets of equations of motion are shown in Fig. \ref{figc200}. 
The results demonstrate good agreement of the predictions for our concerned weakly coupled OMSs, at least within the time period not go beyond all pulse durations considered in the current work.

\section{Conclusion}
\label{conclusion}
In this paper, we study the cooling of OMS with square pulse drive field. Because it is impossible to apply the previous approach of linearizing Eq. (\ref{neq}) 
\cite{pulsetheo1, Liao, Yli} to the cooling with square pulses that change their amplitudes abruptly at heads and tails, we apply a different 
approach in \cite{cooling} for the purpose, and the validity of our treatment is supported by an evidence illustrated in Sec. VI of the current work. The numerical calculations with our linearized equations indicate how to reach a ground state by applying a single square pulse and how to preserve the system to be in the ground ground by using a series of identical pulses. We discuss the relation between the effectiveness and speed of cooling with such square pulses, with which one can design an optimal cooling procedure for a given setup. The cooling with square pulses enjoys more flexibility than the corresponding setup under a CW drive field, so one expects to see more applications of the method in the future experiments.

\section*{Funding}
National Natural Science Foundation of China (NSFC) (11574093); Natural Science Foundation of Fujian Province of China (NSFFPC) (2017J01004); Promotion Program for Young and Middle-aged Teacher in Science and Technology Research of Huaqiao University (PPYMTSTRHU) (ZQN-PY113).

\end{document}